\documentclass[11pt]{article}
\usepackage{amsmath,amssymb,color,graphics,epsfig,cite}

\makeatletter
\@addtoreset{equation}{section}
\makeatother

\usepackage{amsfonts}

\newcommand{\hoch}[1]{$\, ^{#1}$}

\newcommand{\be}{\begin{equation}}
\newcommand{\ee}{\end{equation}}
\newcommand{\bea}{\setlength\arraycolsep{2pt} \begin{eqnarray}}
\newcommand{\eea}{\end{eqnarray}}
\newcommand{\nn}{\nonumber}

\def\crampest{\medmuskip = 1mu plus 1mu minus 1mu}
\def\uncramp{\medmuskip = 4mu plus 2mu minus 4mu}

\def\ft#1#2{{\textstyle{\frac{\scriptstyle #1}{\scriptstyle #2} } }}
\def\fft#1#2{{\frac{#1}{#2}}}

\def\0{{\sst{(0)}}}
\def\1{{\sst{(1)}}}
\def\2{{\sst{(2)}}}
\def\3{{\sst{(3)}}}
\def\4{{\sst{(4)}}}
\def\5{{\sst{(5)}}}
\def\6{{\sst{(6)}}}
\def\7{{\sst{(7)}}}
\def\8{{\sst{(8)}}}
\def\sst#1{{\scriptscriptstyle #1}}

\begin{document}

\begin{flushright}
\hfill{MIFPA-14-32}
\end{flushright}

\vspace{25pt}
\begin{center}
{\large {\bf \Large Black Rings in Supergravity}}

\vspace{20pt}
{\bf 
H. L\"u\hoch{1}, C.N. Pope\hoch{2,3}, Justin F. V\'azquez-Poritz\hoch{4,5} and Zhibai Zhang\hoch{4,5}
}

\vspace{20pt}

\hoch{1}{\it Department of Physics, Beijing Normal University,
Beijing 100875, China}

\vspace{10pt}

\hoch{2} {\it George P. \& Cynthia Woods Mitchell  Institute
for Fundamental Physics and Astronomy,\\
Texas A\&M University, College Station, TX 77843, USA}

\vspace{10pt}

\hoch{3}{\it DAMTP, Centre for Mathematical Sciences,
 Cambridge University,\\  Wilberforce Road, Cambridge CB3 OWA, UK}

\vspace{10pt}

\hoch{4}{\it New York City College of Technology, The City
University of New York,\\ Brooklyn NY 11201, USA}

\vspace{10pt}

\hoch{5}{\it The Graduate School and University Center, The City University of New York,\\ New York NY 10016, USA}

\vspace{40pt}

\underline{ABSTRACT}
\end{center}

We construct black ring solutions in five-dimensional  $U(1)^3$ supergravity which
carry three dipole charges, three electric charges and one 
angular momentum parameter. These solutions are written in a form that is 
sufficiently compact that their global and thermodynamic properties can 
be studied explicitly. In particular, we find that the Smarr formula is obeyed 
regardless of whether or not conical singularities are present, whereas 
the first law of thermodynamics holds only in the absence of conical 
singularities. We also present black ring solutions with three 
background magnetic fields.

\thispagestyle{empty}

\pagebreak

\tableofcontents
\addtocontents{toc}{\protect\setcounter{tocdepth}{2}}


\section{Introduction}

Black holes in five dimensions can have event horizons with $S^1\times S^2$ 
topology, and are known as black rings in order to distinguish them from 
topologically spherical black holes (see \cite{Emparan:2006mm} for a review). 
The existence of black ring solutions demonstrated that black hole 
uniqueness is violated in five dimensions. The neutral black ring in pure 
gravity was presented in \cite{Emparan:2001wn}, and charged black rings in 
supergravity were found in 
\cite{Elvang:2003yy,Elvang:2003mj,Emparan:2004wy,Elvang:2004rt,Gauntlett:2004wh,Bena:2004de,Elvang:2004ds,Gauntlett:2004qy,Elvang:2004xi}.

Various generalizations have been constructed, such as a doubly-spinning 
black ring \cite{Pomeransky:2006bd}, a spherical black hole surrounded by 
a black ring \cite{Elvang:2007rd}, and two concentric orthogonal rotating 
black 
rings \cite{Elvang:2007hs}, as well as approximate solutions describing 
thin, neutral black rings in any dimension $D\ge 5$ in asymptotically 
flat spacetime \cite{Emparan:2007wm} and asymptotically (anti)-de Sitter 
spacetime \cite{Caldarelli:2008pz}. However, while the most general 
supersymmetric solution has been known for quite some 
time \cite{Elvang:2004rt,Bena:2004de,Elvang:2004ds,Gauntlett:2004qy}, the 
general nonextremal solution in five-dimensional $U(1)^3$ supergravity 
has remained elusive until recently \cite{Feldman:2014wxa}. This 
nonextremal black ring has three independent dipole and electric charges, 
along with five additional parameters. If one imposes the condition that 
there are no conical singularities present, then the remaining parameters 
correspond to three electric charges, three dipole charges, 
two angular momenta and the mass.

One solution-generating technique that has been used to construct a number 
of the above black ring solutions is the inverse scattering method, which 
was first adapted to Einstein's equations in \cite{Belinsky:1971nt,Belinsky2}. 
In this paper, we shall make use instead of a solution-generating technique 
that involves performing a dimensional reduction, and which hinges on the 
relation between black rings and the C-metric solution. In particular, 
the neutral black ring can be obtained by lifting the Euclideanized 
Kaluza-Klein C-metric of \cite{Dowker:1993bt} to five dimensions on a 
timelike direction \cite{Chamblin:1996kw,Emparan:2001wn}.

A three-charge generalization of the C-metric has recently been constructed 
in STU gauged supergravity \cite{Lu:2014sza}. In the ungauged limit, 
the Euclidean-signature version of this solution can be lifted on a 
timelike direction to give a black ring solution with three independent 
dipole charges and one non-vanishing rotation parameter. Furthermore, 
three independent electric charges can be generated by repeatedly lifting 
this solution to six dimensions and applying boosts. In a similar manner, 
one can repeatedly lift the C-metric solution to five dimensions and 
apply $SL(2,R)$ transformations to get a black ring solution with 
background magnetic fields, at the expense of altering the asymptotic 
geometry. The large number of parameters, which include dipole charges, 
electric charges and background magnetic field strengths give rise 
to families of black string solutions for which conical singularities 
and Dirac string singularities are completely absent.

The solutions discussed in this paper all have a single non-vanishing 
rotation parameter, as opposed to the black ring presented in 
\cite{Feldman:2014wxa} which has two independent angular momenta. 
In particular, we expect that our black ring with three dipole charges and 
three electric charges arises as a specialization of the one in 
\cite{Feldman:2014wxa}.
The relatively compact form of our solution enables one to verify its 
validity analytically, rather than numerically as done in 
\cite{Feldman:2014wxa}. Moreover, the various properties of the solution, 
its physical and thermodynamic charges and the relations between them 
can be studied explicitly.

This paper is organized as follows. In the next section, we demonstrate 
how black rings with three dipole charges can be obtained from the 
C-metric solutions. We then use solution-generating techniques involving 
dimensional reductions to add three electric charges, as well as 
three background magnetic fields. In section 3, we study the global 
properties of the five and six-dimensional solutions in the Ricci-flat 
limit. In section 4, we perform the global analysis and study the 
thermodynamics of the general black ring solutions. Conclusions are 
presented in section 5. Lastly, in an appendix, we compile the 
dimensional reductions that relate the six, five and four-dimensional 
supergravity theories used for the purposes of solution generating.

\section{Black rings in five-dimensional $U(1)^3$ supergravity}

\subsection{From C-metrics to dipole black rings}

Our starting point is the four-dimensional ungauged supergravity theory 
that is known as the STU model, which can be obtained via a dimensional
reduction from  six-dimensional string theory.  It has 
an $SL(2,R)\times SL(2,R)\times SL(2,R)$ global symmetry, corresponding 
to-- when discretized-- the (S,T,U) duality symmetries of the string theory. 
The theory contains four $U(1)$ vector fields $F_I=dA_I$ and three complex 
scalars $\tau_i=\chi_i + {\rm i} e^{\phi_i}$.   The axions $\chi_i$ 
can be consistently truncated out, provided that one imposes the conditions
\be
F_I\wedge F_J=0\,,\qquad I\ne J\,.\label{fijcons}
\ee
The truncated bosonic Lagrangian is given by
\bea
e_4^{-1} {\cal L}_4 =R_4 -\ft12 \sum_{i=1}^3 (\partial \phi_i)^2
 - \ft14 \sum_{I=1}^4 e^{\vec a_I\cdot \vec \phi} F_I^2\,,\label{stu}
\eea
where the dilaton vectors $\vec a_I$ satisfy
\be\label{a-relation}
\vec a_I \cdot \vec a_J =4\delta_{IJ} -1\,,\qquad
\sum_{I=1}^4 \vec a_I = 0\,.
\ee

Our starting point will be the charged C-metric solutions obtained in 
\cite{Lu:2014sza}, in the ungauged limit $g=0$.  In this case the
solutions can be written as
\bea\label{c-metric}
ds_4^2 &=& \fft{1}{\alpha^2(y-x)^2} \left[ \fft{1}{\sqrt{\tilde U}} \left(G(y) d\tau^2 - \fft{{\cal H}(y)}{G(y)} dy^2\right)
+ \sqrt{\tilde U}\left(\fft{{\cal H}(x)}{G(x)} dx^2 + G(x) d\varphi^2\right)\right]\,,\cr
e^{\vec a_I \cdot \vec\phi} &=& \fft{U_I^2}{\sqrt{\tilde U}}\,,
\qquad F_I=\fft{4Q_I}{h_I(y)^2}\ dy\wedge d\tau\,,\qquad
U_I=\fft{h_I(y)}{h_I(x)}\,,\qquad \tilde U=U_1U_2U_3U_4\,,
\eea
where
\bea\label{notation}
h_I (\xi) &=& 1+ \alpha q_I \xi\,, \qquad {\cal H}(\xi) =\prod_{I=1}^4 h_I(\xi)\,,\cr
G(\xi) &=& {\cal H}(\xi)\Big(b_0 + \sum_{i=1}^4 \fft{16 Q_i^2}{\alpha^2 q_i \prod_{j\ne i} (q_j-q_i)} \fft{1}{h_i(\xi)}\Big)\,.
\eea
The general solution contains 10 parameters: 
$(\alpha, b_0, q_I,Q_I)$ where $I=1,\dots , 4$.  The parameter $b_0$ can 
always be scaled to have one or another of the discrete values $(-1,0,1)$. 
The form of the solution is invariant under
\be
x= \fft{\tilde x}{1 + b \tilde x}\,,\qquad y = \fft{\tilde y}{1 + b \tilde y}\,,\label{xytrans}
\ee
which enables one to set one of the four scalar charges $q_I$ to any value, 
including zero (the STU model has three independent scalars after all).  
In our construction, the parameters $Q_I$ are independent of the scalar 
charges $q_I$.  Therefore, we could set $Q_I=0$ and obtain C-metric 
solutions supported solely by the scalar charges. On the other hand, 
for generic $Q_I$ we cannot set the scalar charges to zero, as one would 
have expected from the theory.  When all of the electric charge parameters 
are equal, namely $Q_I\equiv Q$, we can set all $q_I\equiv q$, the 
three scalar fields all vanish and we recover the charged C-metric solution 
of the Einstein-Maxwell theory. While conical singularities can be 
avoided only for a single point in the parameter space of the dilaton 
C-metric in \cite{Dowker:1993bt}, conical singularities are absent for a 
range of parameters for the present C-metric solutions. We will discuss 
global properties after we lift the solutions to five dimensions.

We shall be lifting the C-metric solution to a dipole black ring solution 
in five dimensions, but before that a number of preliminary steps are
needed. First, we use electromagnetic duality to map the C-metric solution 
to one that has magnetic charges, for which
\be
e^{\vec a_I \cdot \vec\phi}=\fft{\sqrt{\tilde U}}{U_I^2}\,,\qquad F_I = \fft{4Q_I}{h_I(x)^2}\ dx\wedge d\varphi\,.
\ee
Next, we Wick rotate $\tau\rightarrow{\rm i}\psi$ so that the C-metric 
has Euclidean signature. Then we dualize one of the gauge fields so that 
it can play the role of a Kaluza-Klein vector along a timelike direction:
\be
{\cal F}=e^{\vec a_4\cdot \vec \phi} {*F_4} = \fft{4Q_4}{h_4^2}\ dy\wedge d\psi\quad\implies\quad {\cal A}=\fft{4Q_4 y}{h_4(y)} d\psi\,.
\ee
Note that the kinetic term for the corresponding field strength now has 
the ``wrong'' sign in the Lagrangian, which is given by 
(\ref{L-wrong-sign}). However, this is not an issue once we lift to 
five dimensions. Using the reduction ansatz (\ref{4-to-5-timelike}), we 
lift the solution along a timelike direction. The resulting 
five-dimensional solution is given by
\bea\label{ring-solution}
ds_5^2 &=& - \fft{U_4}{(U_1U_2U_3)^{\fft13}} \Big(dt + \fft{4Q_4 y}{h_4(y)} d\psi\Big)^2\cr
&&+\fft{1}{\alpha^2 (x-y)^2} \Big[\fft{1}{U_4(U_1U_2U_3)^{\fft13}}
\Big(-G(y) d\psi^2 - \fft{{\cal H}(y)}{G(y)} dy^2\Big) \cr
&&\qquad\qquad\qquad+
(U_1 U_2 U_3)^{\fft23} \Big(\fft{{\cal H}(x)}{G(x)} dx^2 + G(x) d\varphi^2\Big)\Big]\,,
\cr
F_i &=& \fft{4Q_i}{h_i^2(x)} dx\wedge d\varphi\,,\qquad
e^{\vec b_i \cdot \vec \phi} = \fft{(U_1 U_2 U_3)^{\fft23}}{U_i^2}\,,\qquad
U_i = \fft{h_i(y)}{h_i(x)}\,.\label{3dipole-1}
\eea
This is a black ring solution that has three independent dipole charges. 
The $(x,\varphi)$ subspace corresponds to the $S^2$ part of its $S^2\times S^1$ topology, and the $S^1$ is associated with $\psi$.

In the special case that $Q_i$ and $q_i$ are related as follows:
\be
q_i=-\mu s_i^2\,,\qquad Q_I=\ft14 \mu s_i c_i\,,\qquad
s_i=\sinh\delta_i\,,\qquad c_i=\cosh\delta_i\,,
\ee
we find that
\be
G(\xi)=b_0 {\cal H}(\xi) - \xi^2(1 +\alpha \mu \xi)\,.
\ee
This specialisation enables us to take a Ricci-flat limit of the black 
ring solution, 
by setting $\delta_1=\delta_2=\delta_3=0$, and we shall return to it in 
section \ref{Ricciflatsec}.

\subsection{Adding electric charges}

Using the reduction ansatz (\ref{5-to-6}), we lift the black ring 
solution (\ref{ring-solution}) to six dimensions on the $z$ direction. 
Then we perform a  Lorentz boost $t\rightarrow c_1 t + s_1 z,\ z
\rightarrow c_1 z + s_1 t$,
where $c_i=\cosh\delta_i,\ s_i=\sinh\delta_i$. The reduction back to 
five dimensions along the boosted $z$ direction generates the first 
electric charge, which is associated with the KK vector $A_3$.  Once we 
apply the discrete symmetry to interchange the gauge fields $A_1$ 
and $A_3$ (and rotate the dilatons accordingly), the electric charge is 
now associated with $A_1$.
Next, we repeat the process of lifting to six dimensions, performing a 
boost with parameter $\delta_2$ and reducing back to
five dimensions in order to generate a second electric charge.  Then we 
dualise the 2-form potential $B$ in (\ref{5-to-6}) to
a 1-form potential $A_2$ and use the discrete symmetry again to 
interchange $A_3$ and
$A_2$, so that the second electric charge is now associated with $A_2$.  
Dualising the resulting $A_2$ to a 2-form potential $B$ again, lifting 
to six dimensions and boosting with parameter
$\delta_3$, we again reduce back to five dimensions.  Finally, we 
dualise the
2-form $B$ to a new 1-form $A_2$, thereby arriving at the black ring 
solution
with three independent electric charges.  After making a convenient
transformation of the time coordinate, and a relabelling of the various
functions in the solution, the 3-charge black ring takes the following form:
\be
ds_5^2 = -\fft{U_4}{U^{1/3} (H_1 H_2 H_3)^{2/3}}\,(dt+\omega)^2 +
  (H_1 H_2 H_3)^{1/3}\, ds_4^2\,,\label{3chargemet}
\ee
where
\crampest
\bea
\omega &=& -4 x \Big(\fft{s_1 c_2 c_3 Q_1}{h_1(x)} +
   \fft{c_1 s_2 c_3 Q_2}{h_2(x)} + \fft{c_1 c_2 s_3 Q_3}{h_3(x)} +
        \fft{s_1 s_2 s_3 Q_4}{h_4(x)} \Big) d\varphi\nn\\
&& +
 4 y \Big(\fft{c_1 s_2 s_3 Q_1}{h_1(y)} +
   \fft{s_1 c_2 s_3 Q_2}{h_2(y)} + \fft{s_1 s_2 c_3 Q_3}{h_3(y)} +
        \fft{c_1 c_2 c_3 Q_4}{h_4(y)}\Big) d\psi\,,\nn\\
ds_4^2 &=& \fft1{\alpha^2 (x-y)^2} \left[ \fft{1}{U_4 U^{\fft13}}
\left( -G(y)d\psi^2-\fft{{\cal H}(y)}{G(y)} dy^2\right)+U^{\fft23}
\left( \fft{{\cal H}(x)}{G(x)} dx^2+G(x) d\varphi^2\right)\right]\,
\eea
\uncramp
and
\be
H_i = 1+\left( 1-\fft{U_i^2\, U_4}{U}\right)s_i^2\,,
\qquad U= U_1 U_2 U_3\,.
\ee
The scalar fields are given by
\be
e^{\vec b_i\cdot\vec\phi} = \fft{H_i^2}{U_i^2}\,
   \Big(\fft{U}{H_1 H_2 H_3}\Big)^{2/3}\,,\label{scalarres}
\ee
and the gauge potentials are
\bea
A_1&=&  \fft{s_1 c_1}{H_1}\Big(1-\fft{U_1 U_4}{U_2 U_3}\Big) \, dt +\nn\\
&& +\fft{4x}{H_1}\,\Big( \fft{c_1 c_2 c_3 Q_1}{h_1(x)} +
             \fft{s_1 s_2 c_3 Q_2 U_1 U_4}{U_3 h_2(y)} +
       \fft{s_1 c_2 s_3 Q_3 U_1 U_4}{U_2 h_3(y)} +
\fft{c_1 s_2 s_3 Q_4}{h_4(x)} \Big)\, d\varphi\nn\\
 &&-\fft{4 y}{H_1}\,\Big(\fft{s_1 s_2 s_3 Q_1 U_4}{U_2 U_3 h_1(x)} +
   \fft{c_1 c_2 s_3 Q_2}{h_2(y)} + \fft{c_1 s_2 c_3 Q_3}{h_3(y)} +
     \fft{s_1 c_2 c_3 Q_4 U_1}{U_2 U_3 h_4(x)}\Big)\, d\psi\,,
\label{A1exp}
\eea
with $A_2$ and $A_3$ being obtained from $A_1$ by cycling
the indices 1, 2 and 3 on all quantities appearing in (\ref{A1exp}).

\subsection{Adding background magnetic fields}

We can use a similar solution-generating procedure as in the previous 
section to obtain an Ernst-like generalization of the C-metric solution. 
Using the reduction ansatz (\ref{4-to-5-spacelike}), we lift the 
C-metric solution (\ref{c-metric}) to five dimensions on the 
spacelike direction $z$. Next, we perform the coordinate transformation 
$\varphi\rightarrow\varphi +B z$ and reduce back to four dimensions, where 
we find that $B$ parameterizes the strength of a background magnetic 
field. Applying the discrete symmetry to interchange gauge fields, we keep 
repeating this procedure until we have generated a solution with four 
background magnetic fields, given by
\crampest{
\bea\label{4D-Ernst}
ds_4^2 &=& \fft{1}{\alpha^2 (y-x)^2} \prod_{I=1}^4 \sqrt{\fft{\Lambda_I}{U_I}} \left[ \left( G(y) d\tau^2-\fft{{\cal H}(y)}{G(y)} dy^2\right) +\prod_{I=1}^4 U_I \left(\fft{{\cal H}(x)}{G(x)} dx^2+\fft{G(x)}{\prod_{I=1}^4 \Lambda_I} d\varphi^2\right)\right]\,,\cr
A_I &=& -\fft{1}{B_I\Lambda_I}\left( 1+\fft{4B_IQ_I x}{h_I(x)}\right) d\varphi\,,\qquad
e^{\vec a_I \cdot \vec\phi} = \fft{\prod_{J=1}^4 \sqrt{\Lambda_JU_J}}{\Lambda_I^2 U_I^2}\,,\cr
\Lambda_I &=& \left(1+\fft{4B_IQ_Ix}{h_I(x)}\right)^2+\fft{B_I^2G(x)}{\alpha^2(x-y)^2 U_I^2}\,.
\eea}
The various functions appearing here are defined in (\ref{notation}). 
The C-metric solution (\ref{c-metric}) is recovered for vanishing 
magnetic field parameters $B_I$. Note that one can tune the values of 
the $B_I$ so as to avoid a conical singularity, even if one is present in 
the corresponding C-metric solution (meaning that all $B_I$ are taken to 
zero with all other parameters held fixed). Thus, the background magnetic 
fields associated with $B_I$ play a role analogous to the cosmic string in 
the C-metric itself, by providing the force necessary to accelerate the 
black hole.  (See \cite{cvgiposa} for a discussion of externally magnetised 
charged black hole solutions in STU supergravity.)

We can Wick rotate $\tau\rightarrow {\rm i}\psi$ in the solution (\ref{4D-Ernst}) and then lift it to five dimensions on a timelike direction using the metric ansatz (\ref{4-to-5-timelike}). This yields a dipole black ring with background magnetic fields, given by
%
\bea\label{5D-Ernst}
ds_5^2 &=& \fft{\Lambda^{\fft23}}{\alpha^2 (x-y)^2} \left[\fft{1}{U_4U^{\fft13}}
\left( -G(y) d\psi^2 - \fft{{\cal H}(y)}{G(y)} dy^2\right) +U^{\fft23} \left(\fft{{\cal H}(x)}{G(x)} dx^2 + \fft{G(x)}{\Lambda} d\varphi^2\right)\right]\cr
&-& \fft{U_4}{(\Lambda U)^{\fft13}} \left(dt + \fft{4Q_4 y}{h_4(y)} d\psi\right)^2\,,\cr
A_I &=& -\fft{1}{B_I\Lambda_I}\left( 1+\fft{4B_IQ_I x}{h_I(x)}\right) d\varphi\,,\qquad
e^{\vec b_i \cdot \vec \phi} = \fft{(\Lambda U)^{\fft23}}{(\Lambda_i U_i)^2}\,,\cr
U&\equiv& U_1U_2U_3\,,\qquad \Lambda\equiv \Lambda_1\Lambda_2\Lambda_3\,,
\eea
and $i=1,2,3$. Note that this solution has three independent magnetic field parameters, since in five dimensions one can get rid of $B_4$ by performing 
the reverse of the coordinate transformation that was used to 
generate it in the first place.

\section{Global analysis of Ricci-flat solutions}\label{Ricciflatsec}

\subsection{$D=5$ Ricci-flat metric}

Before turning to the general black ring solution, we shall first study 
the global structure of its Ricci-flat limit, as a warm-up exercise. If 
we set $Q_1=Q_2=Q_3=0=q_1=q_2=q_3$, then the solution 
(\ref{ring-solution}) becomes Ricci-flat and takes the form
\bea
ds_5^2 &=& - \fft{h_4(y)}{h_4(x)} \Big(dt + \fft{4Q_4 y}{h_4(y)} d\psi\Big)^2
+\fft{1}{\alpha^2 (x-y)^2} \Big[\fft{h_4(x)}{h_4(y)}
\Big(-G(y) d\psi^2 - \fft{h_4(y)}{G(y)} dy^2\Big) \cr
&&\qquad\qquad\qquad+
\Big(\fft{h_4(x)}{G(x)} dx^2 + G(x) d\varphi^2\Big)\Big]\,,
\eea
where $h_4(\xi)=1 + \alpha q_4 \xi$. Setting $q_i=0$ can be subtle, 
due to diverging terms in the expression for $G(\xi)$. Instead, we 
substitute the above form of the solution directly into the equations 
of motion in order to obtain the most general solution for $G(\xi)$.
This is given by
\be
G(\xi) = (c_0 + c_1 \xi + c_2 \xi^2) h_4(\xi) - \fft{16 Q_4^2\, \xi^2}{q_4^2}\,.
\ee
We can then shift $c_0,c_1,c_2$ to get rid of the $\xi^2$ factor in 
the $Q_4^2$ term, and shift $\xi$ such that $h_4(\xi)\sim \xi$. This 
results in the expression
\be
G(\xi) = (c_0 + c_1 \xi + c_2 \xi^2) \xi - 
\fft{16Q_4^2}{\alpha^2 q_4^4}\,,\label{newG}
\ee
in which all the parameters of the solution are shown explicitly. Note 
that $G(\xi)$ cannot be written in this form for the general black ring 
solution. The curvature singularities are located at $x=\infty$, 
$y=\infty$ and $h_4(x)=0$, and the asymptotic region is at $x=y$.

Consider the coordinate transformations
\be
x\rightarrow\fft{\tilde x-1}{2q_4^2}\,,
\qquad y\rightarrow\fft{\tilde y-1}{2q_4^2}\,,\qquad
t\rightarrow-\sqrt{a_0}( q_4^2 \tilde t + \psi)\,,
\ee
where $a_0$ is related to $Q_4$ by $Q_4=\ft12\sqrt{a_0} q_4^2$.  Upon 
setting $\alpha=2q_4$, we find that the metric can be written as
\bea
ds_5^2 &=& \fft{1}{(x-y)^2} \Big[\fft{x dx^2}{4G(x)} +
G(x) d\phi^2 - \fft{x dy^2}{4G(y)} - \fft{x G(y) d\psi^2}{y}\Big] -
\fft{a_0y}{x} (dt + y^{-1} d\psi)^2\,,\cr
G(\xi) &=& -a_0 + a_1 \xi + a_2 \xi^2 + a_3 \xi^3\,,\label{d5newring}
\eea
where the parameters $(a_1,a_2,a_3)$ are related to 
$(c_0,c_1,c_2)$ in (\ref{newG}) by
\be
c_0 = \fft{a_1 + a_2 + a_3}{q_4^2}\,,\qquad
c_1=2(a_2 + 2 a_3)\,,\qquad c_2=4a_3 q_4^2\,.
\ee
Then performing the triple Wick rotation
\be
\psi\rightarrow {\rm i} \psi\,,\qquad \phi\rightarrow {\rm i} \phi\,,\qquad t\rightarrow {\rm i} t\,,
\ee
together with $a_0\rightarrow -a_0$, we find that the metric 
(\ref{d5newring}) has the form given by (2.13)-(2.14) in \cite{Lu:2008js}. 
Note that the above triple Wick rotation is equivalent to a double Wick 
rotation on $(x,y)$, which corresponds to taking $a_3\rightarrow - a_3$. 
Since $a_3=-\mu^2<0$ in the black ring analysis of \cite{Lu:2008js}, 
we can let $a_3=\mu^2>0$ and hence
\be
G(\xi)=\mu^2(\xi-\xi_1) (\xi-\xi_2)(\xi-\xi_3)\,,
\qquad a_0=\mu^2 \xi_1\xi_2\xi_3>0\,.
\ee
We shall consider two choices for coordinate ranges.

\subsubsection{Case 1}

We can consider $x\in [\xi_1,\xi_2]$ and $y\in [\xi_2,\xi_3]$ with 
$\xi_3>\xi_2>\xi_1>0$, so that $G(x)\ge 0$ and $G(y)\le 0$.
In contrast to the point $y=0$ for the solution discussed in 
\cite{Lu:2008js}, in the present case there is no ergo-region.
In fact, it turns out that this case describes a smooth soliton which has a 
region with closed timelike curves (CTC's).

\subsubsection{Case 2}

Alternatively, we can consider
\be
G(\xi) = -\mu^2 (\xi-\xi_0) (\xi-\xi_1)(\xi-\xi_2)\,,\qquad a_0=-\mu^2 \xi_0\xi_1\xi_2>0\,,
\ee
with $x\in [\xi_1,\xi_2]$, $y\in [\xi_0,\xi_1]$ and $\xi_0<0<\xi_1<\xi_2$, 
so that $G(x)\ge 0$ and there is a horizon at $y=\xi_0<0$. Then $y=0$ 
constitutes an ergo-region. This is rather similar to the case considered 
in \cite{Lu:2008js}, even though one needs to perform a triple Wick 
rotation in order to relate them.

Following the analysis in \cite{Lu:2008js}, we define
\be
\xi_0=-\eta_0^2\,,\qquad \xi_1\equiv \eta_1^2 < \xi_2\equiv\eta_2^2\,,
\ee
such that $a_0=\mu^2 \eta_0^2\eta_2^2\eta_3^2$ and all $\eta_i$ are 
positive with $\eta_1<\eta_2$. The $\phi$ direction collapses at 
$x=\eta_1^2$ and $x=\eta_2^2$. In order to avoid a conical singularity, 
we must have
\be
\eta_0=\sqrt{\eta_1\eta_2}\,,
\ee
as well as the periodicity condition $\Delta\phi_2=2\pi$, where
\be
\phi_2=\mu^2(\eta_2 - \eta_1) (\eta_1 + \eta_2)^2 \phi\,.
\ee
In order to avoid CTC's at $y=\xi_1$, we take
\be
t\rightarrow t -\fft{\psi}{\eta_1^2}\,.
\ee
Then in order to avoid a conical singularity at $y=\xi_1$ we must 
have $\Delta\phi_1=2\pi$, where
\be
\phi_1=\mu^2(\eta_2 - \eta_1) (\eta_1 + \eta_2)^2\psi\,.
\ee

In order to determine the horizon, we first need to find the asymptotic 
region where $t$ is appropriately defined.  This requires
that
\be
t\rightarrow \fft{t}{\mu (\eta_1\eta_2)^\fft32}\,.
\ee
Making a coordinate transformation
\be
\fft{\sqrt{x-\xi_1}}{x-y} =
\fft{\mu(\eta_2+\eta_1)\sqrt{\eta_2-\eta_1}}{\sqrt{\eta_1}}\,r \cos\theta\,,\qquad
\fft{\sqrt{\xi_1-y}}{x-y} =
\fft{\mu(\eta_2+\eta_1)\sqrt{\eta_2-\eta_1}}{\sqrt{\eta_1}}\,r \sin\theta\,,
\ee
and then letting $r\rightarrow \infty$ yields
\be
ds^2 = - dt^2 + dr^2 + r^2 (d\theta^2 + \cos^2\theta\, d\phi_2^2 + 
\sin^2\theta\, d\phi_1^2)\,.
\ee
On the horizon $y=\xi_0=-\eta_0^2$, the null Killing vector is given by
\be
\ell=\fft{\partial}{\partial_t} + \Omega_\psi \fft{\partial}{\partial_\psi}\,,\qquad
\Omega_\psi=\ft{\mu(\eta_2^2-\eta_1^2) \sqrt{\eta_1}}{\eta_2}\,,
\ee
and the surface gravity is
\be
\kappa=\mu \eta_1(\eta_1+\eta_2)\,.
\ee
Therefore, this solution indeed describes a black ring.  This solution, 
in slightly different coordinates, was first shown to describe a black ring 
in \cite{Emparan:2001wn}.

\subsection{$D=6$ Ricci-flat metric}

Using the reduction ansatz (\ref{5-to-6}) to lift the solution 
(\ref{ring-solution}) to six dimensions yields
\bea\label{6d-solution}
ds_6^2 &=& - \fft{U_4}{\sqrt{U_1U_2}} \Big(dt + \fft{4Q_4 y}{h_4(y)} d\psi\Big)^2
+ \fft{\sqrt{U_1U_2}}{U_3}\Big(dz + \fft{4Q_3 x}{h_3(x)} d\varphi\Big)^2\cr
&&+\fft{1}{\alpha^2 (x-y)^2} \Big[\fft1{U_4\sqrt{U_1U_2}}\Big(-G(y) d\psi^2 - \fft{{\cal H}(y)}{G(y)} dy^2\Big)\cr
&&\qquad\qquad\qquad
+U_3\sqrt{U_1 U_2}\Big(\fft{{\cal H}(x)}{G(x)} dx^2 + G(x) d\varphi^2\Big)\Big]\,,\cr
\hat B &=& \fft{4Q_1x}{h_1(x)} d\varphi\wedge dz + \fft{4Q_2 y}{h_2(y)} dt\wedge d\psi
\,,\qquad e^{\sqrt2\,\phi_1} = \fft{U_2}{U_1}\,.
\eea
As in the five-dimensional case, it is subtle to take the Ricci-flat limit 
of this solution by setting $Q_1=Q_2=0=q_1=q_2$, since information is lost 
upon setting $Q_1=Q_2$ and then $q_1=q_2$. However, this can be remedied 
by first taking the specialization
\be
q_i=-\mu s_i^2\,,\qquad Q_i=\mu s_i c_i\,,\qquad
s_i=\sinh\delta_i\,,\qquad c_i=\cosh\delta_i\,,\qquad i=1,2.
\ee
Then setting $s_i=0$ and renaming the integration constants yields
\be
G(\xi)=(b_0 + b_1 x) h_3(\xi) h_4(\xi) + \fft{16 Q_4^2\, h_3(\xi)}{q_4^3 (q_3-q_4)} + \fft{16 Q_3^2\, h_4(\xi)}{q_3^3 (q_4-q_3)}\,.
\ee
This expression can also be obtained by inserting the form of the 
solution (\ref{6d-solution}) directly into the equations of motion. 
Redefining $(b_0, b_1)$ appropriately results in
\bea
ds_6^2 &=& - U_4 \Big(dt + \fft{4Q_4 y}{h_4(y)} d\psi\Big)^2
+ \fft{1}{U_3}\Big(dz + \fft{4Q_3 x}{h_3(x)} d\varphi\Big)^2\cr
&&+\fft{1}{\alpha^2 (x-y)^2} \Big[\fft{1}{U_4}\Big(-G(y) d\psi^2 - \fft{{\cal H}(y)}{G(y)} dy^2\Big)
+U_3\Big(\fft{{\cal H}(x)}{G(x)} dx^2 + G(x) d\varphi^2\Big)\Big]\,,\cr
G(\xi)&=&(b_0 + b_1 x) h_3(\xi) h_4(\xi) + \fft{16 Q_4^2\, h_3(\xi)\xi^2}{q_4 (q_3-q_4)} + \fft{16 Q_3^2\, h_4(\xi)\xi^2}{q_3 (q_4-q_3)}\,.
\eea

Taking $q_i\rightarrow 1/q_i$ and then shifting $(x,y)$ so that 
$U_4(x,y)=y/x$ enables one to redefine parameters such that the metric 
can be written as
\bea
ds^2 &=& - \fft{y}{x} \Big(dt +Q_4 y^{-1} d\psi\Big)^2 + \fft{x+q_3}{y+q_3}
\Big(dz + \fft{Q_3}{x+q_3} d\phi\Big)\cr
&&+\fft{1}{(x-y)^2}\Big[-\fft{x}{y}\Big(G(y) d\psi^2 + \fft{{\cal H}(y)}{G(y)} dy^2\Big)\cr
&& \qquad+ \fft{y+q_3}{x+q_3}\Big(\fft{{\cal H}(x)}{G(x)} dx^2 + G(x) d\phi^2\Big)
\Big]\,,
\eea
where
\be
G(\xi)=(c_0+c_1 \xi)\xi(\xi+q_3) + q_3^{-1} Q_3^2 \xi -
q_3^{-1} Q_4^2(\xi+q_3)\,,\qquad {\cal H}(\xi)=\fft{\xi(\xi+q_3)}{q_4(q_4+q_3)}\,.
\ee
Curvature singularities occur at $x=\pm\infty$, $y=\pm\infty$, $x=0$ 
and $y=-q_3$. We can express
\be
G(\xi_i)=-\mu^2(\xi-\xi_0)(\xi-\xi_1)(\xi-\xi_2)\,,
\ee
where $\xi_0=-\eta_0^2 <0$ and $\xi_i=\eta_i^2$ ($i=1,2$) with $\eta_2>\eta_1$.  In this parametrization, we have
\be
Q_4=\mu \eta_0\eta_1\eta_2\,,\qquad
Q_3=\mu \sqrt{(\eta_0^2-q_3)(\eta_1^2 + q_3)(\eta_2^2 + q_3)}\,.
\ee

We can consider the coordinate ranges $x\in [\xi_1,\xi_2]$ and 
$y\in [\xi_0,\xi_1]$ so that $G(x)\ge 0$ and $G(y)\le 0$. Next, we make 
coordinate transformations
\bea
t &\rightarrow& t + \fft{2\eta_0\eta_2 \sqrt{\eta_1^2 + q_3}}{(\eta_0^2 +\eta_1^2)(\eta_2^2-\eta_1^2)\mu \sqrt{q_4(q_3+q_4)}} \phi_1\,,\cr
\psi &=& \fft{2\eta_1 \sqrt{\eta_1^2 + q_3}}{(\eta_0^2 +\eta_1^2)(\eta_2^2-\eta_1^2)\mu \sqrt{q_4(q_3+q_4)}}\phi_1\,,\cr
z &\rightarrow & z -\fft{2 \eta_1 \sqrt{(\eta_0^2 - q_3)(\eta_2^2 + q_3)}}{
(\eta_0^2 + \eta_1^2) (\eta_2^2 -\eta_1^2) \mu \sqrt{q_4(
  q_3 + q_4)}}\phi_2\,,\cr
\phi &=&\fft{2 \eta_1 \sqrt{\eta_0^2 - q_3}}{
(\eta_0^2 + \eta_1^2) (\eta_2^2 -\eta_1^2) \mu \sqrt{q_4(
  q_3 + q_4)}}\phi_2\,.
\eea
In the new coordinates, the absence of conical singularities at 
$x=\xi_1$ and $x=\xi_2$ requires that
\be
\Delta\phi_2=2\pi=\Delta\phi_1\,.
\ee
The absence of a conical singularity at $x=\xi_2$ implies that 
$\Delta z=2\pi$. The horizon is located at $y=\xi_0$.  The asymptotic 
region $x=\xi_1=y$ has the geometry (Mink)$_5\times S^1$, where $S^1$ 
corresponds to the $z$ direction. The horizon topology is 
$S^3\times S^1$, where the $S^1$ lies along the $\phi_1$ direction and 
the $S^3$ corresponds to the $(x,\phi_2,z)$ directions.

In order for the asymptotic region to have the geometry (Mink)$_6$ instead 
of (Mink)$_5\times S^1$, we need to decompactify the $z$ direction, by 
taking $Q_3=0$, which corresponds to setting $q_3=\eta_0^2$.  One can then 
attempt to avoid a conical singularity at $x=\xi_{1,2}$ by taking the 
periodicity condition $\Delta\phi_2=2\pi$.  However, it turns that this 
requires that $q_3=0$, and so a conical singularity cannot be avoided.

\section{Global properties and thermodynamics of black rings}

\subsection{Black rings with triple dipole charges}

We shall now study the global structure of black rings that carry three 
dipole charges and obtain the first law of thermodynamics. As it is written 
in (\ref{3dipole-1}), the local solution is over-parameterized.  It is 
advantageous to make the reparametrization
\be
q_i= \fft{1}{\tilde q_4 + \tilde q_i}\,,\qquad
Q_i=-\fft{\widetilde Q_i}{4(\tilde q_4+\tilde q_i)^2 \zeta}\,,\qquad
Q_4=-\fft{\tilde Q_4}{4 \tilde q_4^2 \zeta}\,,\qquad i=1,2,3,\label{repara}
\ee
where $\zeta=\sqrt{\tilde q_4(\tilde q_4+\tilde q_1)(\tilde q_4 + \tilde q_2)(\tilde q_4 + \tilde q_3)}$, followed by the coordinate transformation
\be
x=\tilde x - \tilde q_4\,,\qquad y=\tilde y - \tilde q_4\,,\qquad \varphi=\zeta \tilde\varphi\,,\qquad \psi=\zeta\tilde \psi\,.\label{diffeo}
\ee
We then drop the tilde and redefine $h_i(\xi)$ as
\be
h_i(\xi)\equiv\xi + q_i\,,\qquad i=1,2,3\,,\label{hixi}
\ee
and $h_4(\xi)=\xi$. It turns out that the $U_i$ are given by the same expressions as before, namely
\be
U_i=\fft{h_i(y)}{h_i(x)}\,,\qquad U=U_1 U_2 U_3\,.\label{Uidef}
\ee
The scalar fields and gauge potentials are now given by
\be
e^{\vec b_i \cdot \vec \phi} = \fft{U^{\fft23}}{U_i^2}\,,
\qquad A_i=\fft{Q_i}{h_i(x)} d\varphi\,,\qquad i=1,2,3.
\ee
The metric can now be written as
\bea
ds_5^2&=&U^{-\fft13} \Big( -\fft{y}{x} (dt + Q_4 y^{-1} d\psi)^2 + ds_4^2\Big)\,,\\
ds_4^2&=&\fft{1}{(x-y)^2} \Big[\fft{x}{y} \Big( -G(y) d\psi^2 -\fft{{\cal H}(y)}{G(y)}dy^2\Big)+
U \Big(\fft{{\cal H}(x)}{G(x)} dx^2 + G(x) d\phi^2\Big)\Big]\,,\label{d4met}
\eea
where
\bea
{\cal H}(\xi) &= & \xi\, h_1(\xi)h_2(\xi)h_3(\xi)\,,\cr
G(\xi)&=&{\cal H}(\xi)\Big(b_0 + \fft{Q_1^2}{q_1(q_1-q_2)(q_1-q_3) h_1(\xi)}
+\fft{Q_2^2}{q_2(q_2-q_1)(q_2-q_3)h_2(\xi)}\cr
&&\qquad\qquad +\fft{Q_3^2}{q_3(q_3-q_1)(q_3-q_2)h_3(\xi)} -
\fft{Q_4^2}{q_1 q_2 q_3\xi}\Big)\,.
\eea
An advantage of this parameterization is that $q_4$ drops out from the 
solution completely.  We have also set $\alpha=1$ without loss of generality.

We are now in the position to study the global structure of the solution.  
The asymptotic region is located at $x=y$. Curvature singularities arise 
when either $h_i(x)$ or $h_i(y)$ vanishes or when either $x$ or $y$ 
diverges.  Thus, we should arrange that the ranges of the coordinate 
$x$ and $y$ are confined to intervals that are specified by adjacent 
roots of $G(\xi)$, within which the $h_i(\xi)$ are non-vanishing.  It is 
therefore more convenient to express $G(\xi)$ in terms of four roots, namely
\be
G(\xi) = - \mu^2 (\xi-\xi_1)(\xi-\xi_2)(\xi-\xi_3)(\xi-\xi_4)\,.\label{4roots}
\ee
Here we let $b_0=-\mu^2 <0$.  The charge parameters $Q_i$ and $Q_4$ can be 
expressed in terms of $\mu$ and the four roots as
\bea
Q_i^2 &=& \mu^2 (q_i + \xi_1)(q_i + \xi_2)(q_i + \xi_3)(q_i + \xi_4)\ge 0\,,\cr
Q_4^2 &=& \mu^2 \xi_1\xi_2\xi_3\xi_4\ge 0\,.
\eea

As with the Ricci-flat metrics discussed in the previous section, the
sugns of $G(x)\ge 0$ and $G(y)\le 0$ should be such that the metric has 
the proper signature.  Without loss of generality, we consider 
$x\in [\xi_3,\xi_4]$ and $y\in [\xi_2,\xi_3]$.  

    The reality of $Q_4$ leads 
to two possibilities. The first one is where $0<\xi_1<\xi_2<\xi_3<\xi_4$.  
After a careful analysis, we find that the absence of naked curvature 
power-law singularities, together with the reality condition on $Q_i$, 
implies that the solution has an unavoidable naked conical singularity.  
This conclusion may not be too surprising, given that we would otherwise 
have a rotating black hole without an ergo-region. The natural location of 
the ergo-region is $y=0$, which leads to the second case, for which
\be
\xi_1< \xi_2 <0<\xi_3<\xi_4\,.\label{xisize}
\ee
This ensures that the ergo-region at $y=0$ lies in the range 
$y\in[\xi_2,\xi_3]$.  The metric has no naked power-law curvature 
singularities provided that
\bea
&&h_i(x)>0\qquad \hbox{for} \qquad x\in [\xi_3,\xi_4]\,,\cr
&&h_i(y)>0\qquad \hbox{for} \qquad y\in [\xi_2,\xi_3]\,.\label{powerlawcond}
\eea
Now we ensure that the solution does not have conical singularities. 
The null Killing vectors with unit Euclidean surface gravity at 
$x=\xi_3$ and $x=\xi_4$ are given by
\be
\ell_{x=\xi_3} = \alpha_3 \partial_{\phi}\,,\qquad
\ell_{x=\xi_4}=\alpha_4 \partial_{\phi}\,,
\ee
where
\be
\alpha_{3} = \fft{2\sqrt{\xi_3(\xi_3+q_1)(\xi_3+q_2)(\xi_3+q_3)}}{
\mu^2(\xi_3-\xi_1)(\xi_3-\xi_2)(\xi_4-\xi_3)}\,,\quad
\alpha_{4} =\fft{2\sqrt{\xi_4(\xi_4+q_1)(\xi_4+q_2)(\xi_4+q_3)}}{
\mu^2(\xi_4-\xi_1)(\xi_4-\xi_2)(\xi_4-\xi_3)}\,.
\ee
The absence of conical singularities then requires that $\alpha_3=\alpha_4$. 
This condition and (\ref{powerlawcond}) can be shown to be simultaneously 
satisfied for the appropriate choice of parameters. As an example, 
consider $\xi_1=-2$, $\xi_2=-1$, $\xi_3=1$ and $\xi_4=2$, for which the 
absence of conical singularities requires
\be
\fft{(q_1+2)(q_2+2)(q_3+2)}{2(q_1+1)(q_2+1)(q_3+1)}=1\,.
\ee
This can be solved, for example, with $q_1=2$, $q_2=3$ and $q_3=4$.  All 
the $q_i\ge-2$, so that $Q_i\ge 0$, $h_i(x)>0$ and $h_i(y)>0$.  Thus, 
there are no singularities in the region of interest.

Once we have established that $\alpha_3=\alpha_4$ and that the charge 
parameters $Q_i$ are real, there are no further conditions on the 
parameters. The absence of a conical singularity at $y=\xi_3$ tells us 
the appropriate period for the coordinate $\psi$, and the horizon 
condition gives the temperature and entropy. However, the algebraic 
expression for $\alpha_3=\alpha_4$ is rather complicated to solve, which 
means that quantities such as mass, charges and temperature can be quite 
complicated.

To proceed, it is convenient to rewrite the roots as
\be
\xi_1=-\eta_1^2\,,\qquad \xi_2=-\eta_2^2\,,\qquad
\xi_3=\eta_3^2\,,\qquad \xi_4=\eta_4^2\,,
\ee
with $\eta_1>\eta_2>0$ and $\eta_4>\eta_3>0$.  The avoidance of naked 
conical singularities requires that
\be
\fft{\eta_4(\eta_1^2 + \eta_3^2)(\eta_2^2 + \eta_3^2)\sqrt{(q_1+\eta_4^2)(q_2+\eta_4^2)(q_3+\eta_4^2)}}{
\eta_3(\eta_1^2 + \eta_4^2)(\eta_2^2 + \eta_4^2)\sqrt{(q_1+\eta_3^2)(q_2+\eta_3^2)(q_3+\eta_3^2)}}=1\,.\label{ringcond}
\ee
We define a new set of coordinates, given by
\bea
&&\phi=a\,\phi_2\,,\qquad \psi=a\,\phi_1\,,\qquad
t\rightarrow t -\fft{Q_4a}{\eta_3^2} \phi_1\,,\cr
&&a=\fft{2\eta_3 \sqrt{(q_1+\eta_3^2)(q_2+\eta_3^2)(q_3+\eta_3^2)}}{
\mu^2 (\eta_1^2 + \eta_3^2)(\eta_2^2 + \eta_3^2)(\eta_4^2-\eta_3^2)}\,.
\eea
Then the periods of $\phi_1$ and $\phi_2$ are both $2\pi$. Note that the 
shift in $t$ ensures that only the spatial coordinate $\phi_1$ collapses 
to zero size at $y=\xi_3$, and hence we avoid naked closed timelike 
curves (CTC's).

On the horizon at $y=\xi_2$, the null Killing vector is
\be
\ell=\partial_t + \Omega_+ \partial_{\phi_1}\,,\qquad
\Omega_+=\fft{\eta_2^2\eta_3^2}{a (\eta_2^2 + \eta_3^2) Q_4}\,.
\ee
The (Euclidean) surface gravity is given by
\be
\kappa^2 = -\fft{\mu^4
\eta_2^2\eta_3^2(\eta_1^2-\eta_2^2)^2(\eta_2^2+\eta_4^2)^2}{4
(q_1-\eta_1^2)(q_2-\eta_2^2)(q_3-\eta_3^2) Q_4^2}\,.
\ee
For the solution to describe a black object with a horizon, we must 
have $\kappa^2<0$.  It is worth checking that this condition can indeed 
be satisfied. As an example, we take
\be
\xi_1=-2\,,\qquad \xi_2=-1\,,\qquad \xi_3=1\,,\qquad \xi_4=4\,.
\ee
An acceptable set of $q_i$ with $i=1,2,3$ must satisfy (\ref{ringcond}) 
and all have $q_i>-2$. Such solutions do in fact exist; for instance,
\be
q_1=\ft73\,,\qquad q_2=\ft94\,,\qquad q_3=\ft{68971}{27389}\,.
\ee
For this choice of parameters the $Q_i$ are real, $h_i(x)$ and $h_i(y)$ 
are positive definite in the regions of concern, $\kappa^2<0$ and $\chi$ 
is real. Thus, a well-behaved black ring exists.  The temperature and 
entropy are given by
\be
T=\fft{\kappa}{2\pi}\,,\qquad
S=\fft{\pi^2 a^2Q_4(\eta_4^2-\eta_3^2) \sqrt{(q_1-\eta_2^2)(q_2-\eta_2^2)(q_3-\eta_2^2)}}{
\eta_2\eta_3^3(\eta_2^2 + \eta_4^2)}\,.
\ee
When $\eta_1=\eta_2$, the temperature vanishes, corresponding to the 
extremal limit.

The asymptotic region is located at $x=\xi_3=y$.  To see this, we make 
the coordinate transformation
\bea
&&\fft{\sqrt{x-\xi_3}}{x-y} =
b\,r \cos\theta\,,\qquad
\fft{\sqrt{\xi_3-y}}{x-y} =
b\,r \sin\theta\,,\cr
&&b^2 = \fft{\mu^2(\eta_1^2+\eta_3^2)(\eta_2^2+\eta_3^2)(\eta_4^2-\eta_3^2)}{4\eta_3^2(\eta_3^2 + q_1)(\eta_3^2 + q_2)
(\eta_3^2 + q_3)}\,,
\eea
and then we take $r\rightarrow \infty$. The asymptotic geometry is 
five-dimensional Minkowski spacetime, with the metric written as
\be
ds^2_5= -dt^2 + dr^2 + r^2 (d\theta^2 + \sin^2\theta d\phi_1^2 +
\cos^2\theta d\phi_2^2)\,.
\ee
From the asymptotic falloffs of the metric, we can read off the ADM mass 
and the angular momenta as
\be
M = \fft{\pi}{8b^2}\Big(\fft{3}{\eta_3^2} - \fft{1}{\eta_3^2+q_1} -
\fft{1}{\eta_3^2+q_2} - \fft1{\eta_3^2+q_3}\Big)\,,
\ee
and
\be
J_{\phi_1} = \fft{\pi a Q_4}{4b^2\eta_3^4}\,,\qquad
J_{\phi_2}=0\,,
\ee
respectively. The dipole charges are given by
\be
D_i = \ft{1}{8} \int F_i = \ft14 \pi a Q_i \Big(\fft{1}{q_i + \eta_3^2} -
\fft{1}{q_i + \eta_4^2}\Big)\,.
\ee
The only quantities left to determine are the dipole potentials 
$\Phi_{D_i}$, which requires the dualization of $A_\mu^i$ to 
$B_{\mu\nu}^i$.  Since $F_i\wedge F_j=0$, we do not need to worry about 
the $FFA$ term in the Lagrangian when performing the dualization.  We find
\be
e^{\vec b_i\cdot\vec \phi} {*F_i} = \fft{a Q_i}{ (y+ q_i)^2} dt \wedge d\phi_1\wedge dy\,,\qquad B_i = -\fft{a Q_i}{y + q_i} dt \wedge d\phi_1\,.
\ee
The 2-form potential difference between the horizon and the asymptotic 
region is then given by
\be
\Phi_{D_i} = a Q_i\Big(\fft{1}{q_i-\eta_2^2}-\fft{1}{q_i+\eta_3^2}\Big)\,.
\ee

Having obtained all of the thermodynamic quantities, it is straightforward 
to verify that the first law of the thermodynamics,
\be
dM = T dS + \Omega_+ dJ_{\phi_1} + \sum_{i=1}^3\Phi_{D_i} dD_i\,,
\ee
is obeyed.
The Smarr formula is given by
\be
M=\ft32 TS + \ft32 \Omega_{\phi_1} J_{\phi_1} + \sum_{i=1}^3 \ft12 \Phi_{D_i} dD_i\,.
\ee
Interestingly enough, the Smarr formula is actually valid even without 
imposing the condition (\ref{ringcond}) that ensures the absence of 
naked conical singularities.  The solution involves five non-trivial 
parameters associated with the mass, one non-vanishing angular momentum, 
and three dipole charges.  This solution is over-parameterized by 
two trivial parameters. While the $\mu$ parameter can be absorbed by 
a ``trombone'' scaling of the metric and other fields, the second 
extra parameter is more subtle.  Although the five-dimensional theory 
has only two scalars, the solution has three scalar charges $q_i$, 
one of which is therefore trivial.

The horizon geometry has the metric
\bea
ds_3^2 &=& \Big(\fft{(x+q_1)(x+q_2)(x+q_3)}{(q_1-\eta_2^2)(q_2-\eta_2^2)(q_3-\eta_2^2)^2}
\Big)^{\fft13}\Big[\fft{(q_1-\eta_2^2)(q_2-\eta_2^2)(q_3-\eta_3^2)\, x}
{\mu^2(x+\eta_1^2)(x+\eta_2^2)^3(\eta_4^2-x)(x-\eta_3^2)} dx^2\cr
&&+\fft{a^2\mu^2(q_1-\eta_2^2)(q_2-\eta_2^2)(q_3-\eta_3^2)(x+\eta_1^2)
(\eta_4^2-x)(x-\eta_3^2)}{(x+q_1)(x+q_2)(x+q_3)(x+\eta_2^2)} d\phi_2^2\cr
&&+ \fft{a^2 Q_4^2(\eta_2^2 + \eta_3^2)^2}{\eta_2^2 \eta_3^4\, x} d\phi_1^2\Big]\,.
\eea
Since $x\in[\eta_3^2,\eta_4^2]$, it is clear that the horizon topology is 
$S^2\times S^1$. With non-vanishing dipole charges, the black ring 
solution does not have a limit with a spherical horizon.  

\subsection{Electrically-charged black rings and naked CTC's}

We shall now study the global structure of the black ring solutions 
with three dipole charges and three electric charges, which was 
obtained in section 2.2.  As before, it is convenient to make the 
reparametrizations (\ref{repara}) and the coordinate 
transformations (\ref{diffeo}).  Furthermore, we make a redefinition of the
time coordinate,
\bea
t&=&\tilde t - \Big(\fft{s_1 c_2 c_3 \widetilde Q_1}{\tilde q_4 + \tilde q_1}
+ \fft{c_1 s_2 c_3 \widetilde Q_2}{\tilde q_4 + \tilde q_2} +
\fft{c_1 c_2 s_3 \widetilde Q_3}{\tilde q_4 + \tilde q_3} +
\fft{s_1 s_2 s_3 \widetilde Q_4}{\tilde q_4}\Big) \tilde \varphi\cr
&&\qquad\,\, +\Big(\fft{c_1 s_2 s_3 \widetilde Q_1}{\tilde q_4 + \tilde q_1}
+ \fft{s_1 c_2 s_3 \widetilde Q_2}{\tilde q_4 + \tilde q_2} +
\fft{s_1 s_2 c_3 \widetilde Q_3}{\tilde q_4 + \tilde q_3} +
\fft{c_1 c_2 c_3 \widetilde Q_4}{\tilde q_4}\Big) \tilde \psi\,.
\eea
We may now  drop the tilde, and redefine $h_i(\xi)$ as in (\ref{hixi}). 
The functions $H_i$ take the same form, namely
\be
H_i = 1 + \Big(1 - \fft{yU_i^2}{x U}\Big) s_i^2\,,
\ee
where $U_i$ and $U$ are given by (\ref{Uidef}). The scalar fields retain 
the same form given by (\ref{scalarres}).  The metric is now given by
\be
ds_5^2 = \fft{1}{U^{1/3} (H_1 H_2 H_3)^{2/3}}\Big(-\fft{y}{x}\,(dt+\omega)^2 +
  H_1 H_2 H_3\, ds_4^2\Big)\,,\label{3chargemet2}
\ee
where $ds_4^2$ is precisely the same as (\ref{d4met}) and $\omega$ is
\bea
\omega &=& -\Big(\fft{s_1 c_2 c_3 Q_1}{h_1(x)} +
   \fft{c_1 s_2 c_3 Q_2}{h_2(x)} + \fft{c_1 c_2 s_3 Q_3}{h_3(x)} +
        \fft{s_1 s_2 s_3 Q_4}{x} \Big) d\varphi\cr
&& + \Big(\fft{c_1 s_2 s_3 Q_1}{h_1(y)} +
   \fft{s_1 c_2 s_3 Q_2}{h_2(y)} + \fft{s_1 s_2 c_3 Q_3}{h_3(y)} +
        \fft{c_1 c_2 c_3 Q_4}{y}\Big) d\psi\,.
\eea
The gauge potentials are
\bea
A_1&=&  \fft{s_1 c_1}{H_1}\Big(1-\fft{U_1 U_4}{U_2 U_3}\Big) \, dt +\nn\\
&& +\fft{1}{H_1}\,\Big( \fft{c_1 c_2 c_3 Q_1}{h_1(x)} +
             \fft{s_1 s_2 c_3 Q_2 U_1 U_4}{U_3 h_2(y)} +
       \fft{s_1 c_2 s_3 Q_3 U_1 U_4}{U_2 h_3(y)} +
\fft{c_1 s_2 s_3 Q_4}{x} \Big)\, d\varphi\nn\\
 &&-\fft{1}{H_1}\,\Big(\fft{s_1 s_2 s_3 Q_1 U_4}{U_2 U_3 h_1(x)} +
   \fft{c_1 c_2 s_3 Q_2}{h_2(y)} + \fft{c_1 s_2 c_3 Q_3}{h_3(y)} +
     \fft{s_1 c_2 c_3 Q_4 U_1}{U_2 U_3\, x}\Big)\, d\psi\,,
\label{A1exp1}
\eea
with $A_2$ and $A_3$ being obtained from $A_1$ by cycling
the indices 1, 2 and 3 on all quantities appearing in (\ref{A1exp1}).  Note 
that, aside from the coordinate transformations and reparametrizations, 
the $A_i$ obtained above are related to those in (\ref{A1exp}) by 
gauge transformations.

It is of interest to note that the dipole charges are magnetic and are 
associated with the $(x,\varphi)$ directions. Adding the electric charges 
has the effect of producing angular momentum in the $\varphi$ direction 
as well.  However, this also has the undesirable side effect of 
producing naked CTC's. To see this explicitly, it is useful to note 
that $ds_4^2$ is identical to that in the previous subsection and $G(\xi)$ 
can be expressed by (\ref{4roots}) with (\ref{xisize}). The null Killing 
vector associated with the collapsing circles at $x=\xi_3$ and $x-\xi_4$ are
given by
\be
\ell_{x=\xi_3} = \alpha_3 \partial_{\phi} +\beta_3 \partial_t\,,\qquad
\ell_{x=\xi_4}=\alpha_4 \partial_{\phi} + \beta_4\partial_t\,.
\ee
The absence of CTC's requires that 
\be\fft{\beta_3}{\alpha_3}=\fft{\beta_4}{\alpha_4}\,,\label{ctccon}
\ee
in which case we can shift $t\rightarrow t + \gamma \phi$, for an 
appropriate constant $\gamma$, such that the null Killing vectors do not 
involve the newly-defined time. For black rings with dipole charges but 
no electric charges, $\beta_3=0=\beta_4$ and hence there are no naked CTC's. 
By contrast, we find that for black rings with both dipole and electric 
charges, the condition (\ref{ctccon}) cannot be satisfied and the solutions 
have naked CTC's. This parallels the situation for the five-dimensional 
black hole solutions studied in \cite{Lu:2008js}, for which naked CTC's
appeared when the electric charges were nonvanishing. Turning on two independent angular momenta might result in black ring solutions with dipole and electric charges but without naked CTC's.

\subsection{In background magnetic fields}

We shall now consider some of the global properties of the solution with
external magnetic fields, with 
the metric (\ref{5D-Ernst}). We use the same parameterizations as in 
section 4.1, with $G(\xi)$ given by (\ref{4roots}) and with the coordinate 
ranges $x\in [\xi_3,\xi_4]$ and $y\in [\xi_2,\xi_3]$. In order to avoid 
a conical singularity in the $(x,\phi)$ subspace, we must have
\be\label{B-condition}
\prod_{i=1}^3\left(\fft{\xi_4+q_i+B_iQ_i(\xi_4-q_4)}{\xi_3+q_i+B_iQ_i(\xi_3-q_4)}\ \sqrt{\fft{\xi_3+q_i}{\xi_4+q_i}}\right)=\fft{(\xi_4-\xi_1)(\xi_4-\xi_2)}{(\xi_3-\xi_1)(\xi_3-\xi_2)}\ \sqrt{\fft{\xi_3}{\xi_4}}\,,
\ee
along with the appropriate periodicity for $\psi$. 

First consider 
the case where $0<\xi_1<\xi_2<\xi_3<\xi_4$. While it is not possible to 
satisfy the condition (\ref{B-condition}) and the reality condition on 
the charge parameters $Q_i$ at the same time for vanishing $B_i$, both 
conditions can be satisfied simultaneously when the $B_i$ are turned on. 
A sample solution is given by
\bea
B_1 &=& -1\,,\quad B_2=-2\,,\quad B_3=-3\,,\quad \xi_1=1\,,\quad \xi_2=2\,,\quad \xi_3=3\,,\quad \xi_4=4\,,\cr
\mu &=& 0.2\,,\quad q_2=1\,,\quad q_3=1.5\,,\quad q_4=2.5\,,\quad q_1=2.09789\,.
\eea
Thus, at the expense of altering the asymptotic geometry, turning \
on background magnetic fields can have the effect of removing conical 
singularities for black rings, in much the same way as it does for 
Ernst solutions \cite{ernst,Dowker:1993bt}.

Background magnetic fields also enable us to have multiple branches 
of solutions. For instance, in the case where $\xi_1< \xi_2 <0<\xi_3<\xi_4$ 
with vanishing $B_i$, a sample solution is given by
\be
\xi_1=-2\,,\quad \xi_2=-1\,,\quad \xi_3=1\,,\quad \xi_4=2\,,\quad q_1=2\,,\quad q_2=3\,,\quad q_3=4\,.
\ee
Note that if only $q_1$ were to be left unspecified then the condition 
(\ref{B-condition}) would uniquely determine its value in terms of the 
other parameters. On the other hand, for nonvanishing $B_i$, sample 
solutions are given by
\bea
B_1 &=& 1\,,\quad B_2=2\,,\quad B_3=3\,,\quad \xi_1=-2\,,\quad \xi_2=-1\,,\quad \xi_3=1\,,\quad \xi_4=2\,,\cr
\mu &=& 1\,,\quad q_2=2\,,\quad q_3=3\,,\quad q_4=4\,,\quad q_1=2.03367\ \mbox{or}\ 0.975202\,.
\eea
Due to the presence of the $B_i$, using the condition (\ref{B-condition}) 
to solve for $q_1$ in terms of the other specified parameters yields 
two different solutions, both of which satisfy the reality condition 
on the charge parameters $Q_i$.

\section{Conclusions}

We have constructed black ring solutions in five-dimensional  
$U(1)^3$ supergravity,  carrying three independent dipole charges, 
three electric charges and one non-vanishing angular momentum. We have 
also presented black ring solutions with three background magnetic fields. 
These various solutions have been obtained by lifting the Euclidean 
C-metric solution of four-dimensional ungauged STU 
supergravity \cite{Lu:2014sza} to five dimensions on a timelike direction, 
and then using solution-generating techniques involving dimensional 
reductions to add electric charges or background magnetic fields. We find 
that adding the electric charges gives rise to black rings with naked CTC's.

We expect that the solutions without the background magnetic fields should 
arise as special cases of the black ring solutions obtained in 
\cite{Feldman:2014wxa} if 
one of the angular momenta is set to zero. We have expressed this 
specialization in a form that is sufficiently compact that its various 
physical properties can be investigated explicitly. In particular, its 
global structure has been analyzed and the conditions determined in order 
for conical singularities and Dirac string singularities to be absent.
Expressions for its mass, dipole charges, electric charges and angular 
momentum have been obtained, as well as the temperature and entropy. 
Moreover, we have analyzed the thermodynamics, finding that the Smarr 
formula is obeyed regardless of whether or not conical singularities 
are present.  By contrast, the first law of thermodynamics is obeyed
only in those cases where conical singularities are absent.

The four-dimensional Ernst-like generalization of the C-metric solution 
obtained in this paper can be Wick rotated to a Euclidean instanton that 
describes the pair creation of black holes in magnetic fields. This 
generalizes the one-parameter family of instantons in 
\cite{Garfinkle:1990eq,Garfinkle:1993xk,Dowker:1993bt} to multiple 
parameters. This substantially enhances the families of explicit examples 
for the creation of maximally entangled black holes, for which it has 
recently
been proposed that they may be connected by some kind of 
Einstein-Rosen bridge \cite{Maldacena:2013xja}.

The exact time-dependent C-metric solution constructed in 
\cite{Lu:2014ida} can be embedded in STU supergravity which, in the 
ungauged limit, can be lifted to five dimensions. It would be interesting 
to analyze the global structures of these five-dimensional solutions, 
especially with the prospect of finding time-dependent black rings.

\section*{Acknowledgements}

We are grateful to Mboyo Esole for helpful conversations. The work of C.N.P. 
was supported in part by DOE grant  DE-FG02-13ER42020; the work of H.L. 
was supported in part by NSFC grants 11175269, 11475024 and 11235003; 
the work of J.F.V.P. was supported in part by a PSC-CUNY Award.

\appendix

\section{Dimensional reductions}

We present the Kaluza-Klein dimensional reductions that have been used to 
relate the four-dimensional C-metrics with the five-dimensional black 
ring solutions, as well as for the purposes of generating electric charges 
and background magnetic fields. We start with the $D=6$ theory whose 
Lagrangian is given by
\be
e_6^{-1}{\cal L}_6 = R_6 - \ft12(\partial\phi_1)^2 - \ft1{12} e^{\sqrt2\phi_1} \hat H^2\,,
\ee
where $\hat H=d\hat B$. Consider the reduction ansatz
\bea\label{5-to-6}
ds_6^2 &=& e^{-\fft{1}{\sqrt6}\phi_2} ds_5^2 + e^{\fft{3}{\sqrt6}\phi_2} (dz + A_3)^2\,,\cr
\hat B &=& B + A_1\wedge dz\,,
\eea
where
\be
F_2=e^{-b_2\cdot \vec \phi} {*H}\,,\qquad H=dB-A_1\wedge A_3\,.
\ee
This yields the Lagrangian for the bosonic sector of 
five-dimensional $U(1)^3$ supergravity, given by
\be
e_5^{-1} {\cal L}_5 = R_5- \ft12 (\partial\vec\phi)^2 - \fft14\sum_{i=1}^3 e^{\vec b_i \cdot\vec \phi} \hat F_i^2 + {\cal L}_{FFA}\,,
\ee
where $\vec \phi=(\phi_1,\phi_2)$ and the dilaton vectors $\vec b_i$ are given by
\be
\vec b_1=(\sqrt2, -\ft2{\sqrt6})\,,\qquad \vec b_2=(-\sqrt2, -\ft2{\sqrt6})\,,\qquad
\vec b_3=(0,\ft4{\sqrt6})\,,
\ee
which obey
\be
\vec b_i \cdot \vec b_j = 4 \delta_{ij} - \ft{4}{3}\,,\qquad
\sum_{i=1}^3 \vec b_i=0\,.
\ee

Next, we perform a dimensional reduction to the four-dimensional 
$U(1)^4$ theory.\footnote{Note that we are truncating out the three axions
that would arise in the reductions of the three five-dimensional
gauge fields.  This truncation is consistent with the equations of
motion provided that we restrict our attention to solutions for which
$F^i\wedge F^j=0$.}  Reducing on a spacelike direction with the metric ansatz
\be\label{4-to-5-spacelike}
ds^2_5 = e^{-\fft{1}{\sqrt3}\phi_3} ds_4^2 + e^{\fft{2}{\sqrt3}\phi_3} (dz+A_4)^2\,,
\ee
yields
\be\label{symmetric-form-4dtheory}
e_4^{-1} {\cal L}_4 = R_4- \ft12 (\partial\vec\phi)^2 -\fft14\sum_{i=1}^4 e^{\vec a_i \cdot\vec \phi} \hat F_i^2 \,,
\ee
where we are now taking $\vec \phi=(\phi_1,\phi_2,\phi_3)$. The dilaton vectors $\vec a_i$ are given by
\be
\vec a_i=(\vec b_i, \ft{1}{\sqrt3} )\,,\qquad
\vec a_4 = (0,0,-\sqrt3)\,,
\ee
and they satisfy the conditions in (\ref{a-relation}).

Alternatively, we can reduce on the timelike direction with the metric ansatz
\be\label{4-to-5-timelike}
ds^2_5 = e^{-\fft{1}{\sqrt3}\phi_3} ds_4^2 - e^{\fft{2}{\sqrt3}\phi_3} (dt - {\cal A})^2\,,
\ee
which gives
\be\label{L-wrong-sign}
e_4^{-1} {\cal L}_4 = R_4- \ft12 (\partial\vec\phi)^2 -\fft14\sum_{i=1}^3 e^{\vec a_i \cdot\vec \phi} \hat F_i^2 + \fft14 e^{-\vec a_4\cdot \vec\phi} {\cal F}^2\,.
\ee
If we perform a Hodge dualization on the Kaluza-Klein vector, namely
\be
e^{-\vec a_4\cdot \vec \phi} {*{\cal F}} = F_4\,,
\ee
then the kinetic term changes sign and the four-dimensional Lagrangian 
can be expressed in the more symmetric form given by 
(\ref{symmetric-form-4dtheory}).

\end{document}